\documentclass[figures]{epl}
\usepackage{graphics,amssymb}

\def\lett{\cite{BaSy02}}

\begin{document}

\title{Comment on ``On the problem of initial conditions in
  cosmological N-body simulations'' (EpL {\bf 57}, 322)}
\shorttitle{Comment}

\author{A. Dom{\'\i}nguez\inst{1}
  \and A. Knebe\inst{2}
}

\institute{
  \inst{1} Max-Planck-Institute f\"ur Metallforschung, Heisenbergstr.~3, D-70569 Stuttgart, 
  FRG \\
  \inst{2} Centre for Astrophysics \& Supercomputing, Swinburne
  University, \\ Mail \#31, P.O. Box 218, Hawthorn, Victoria, 3122,
  Australia}

\shortauthor{Dom{\'\i}nguez \& Knebe}

\maketitle

In the Letter \lett, the initial conditions (IC's) of cosmological
N-body simulations by the {\it Virgo Consortium} \cite{JFPT98} are
analyzed and it is concluded that the density fluctuations are rather
different from the desired ones. 
We have repeated the analysis of the IC's using our own code and the
code provided by the authors of \lett, obtaining results that disprove
the criticisms.

\begin{figure}[h]
  \resizebox{.5\hsize}{!}{\includegraphics{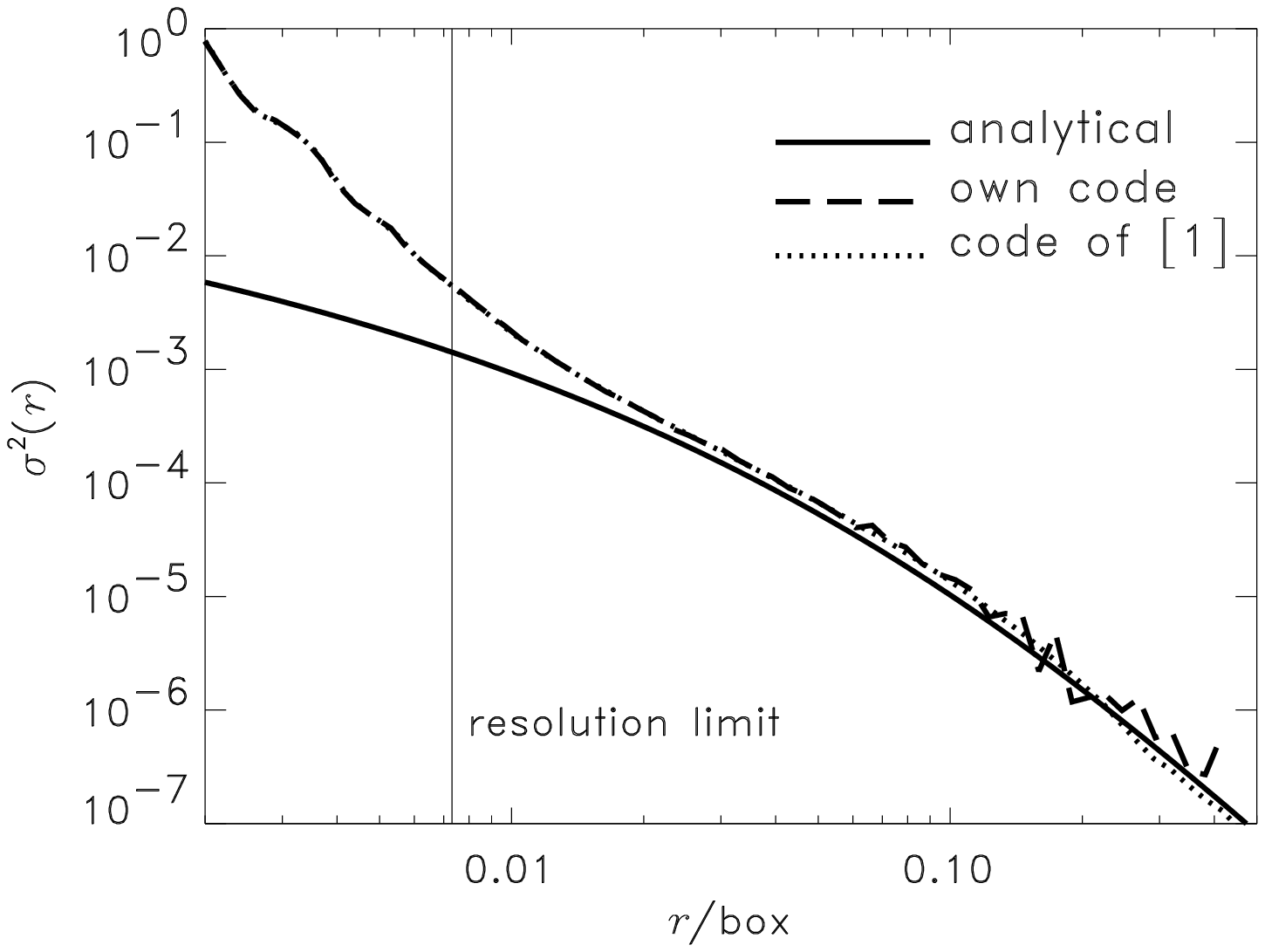}}
  \resizebox{.5\hsize}{!}{\includegraphics{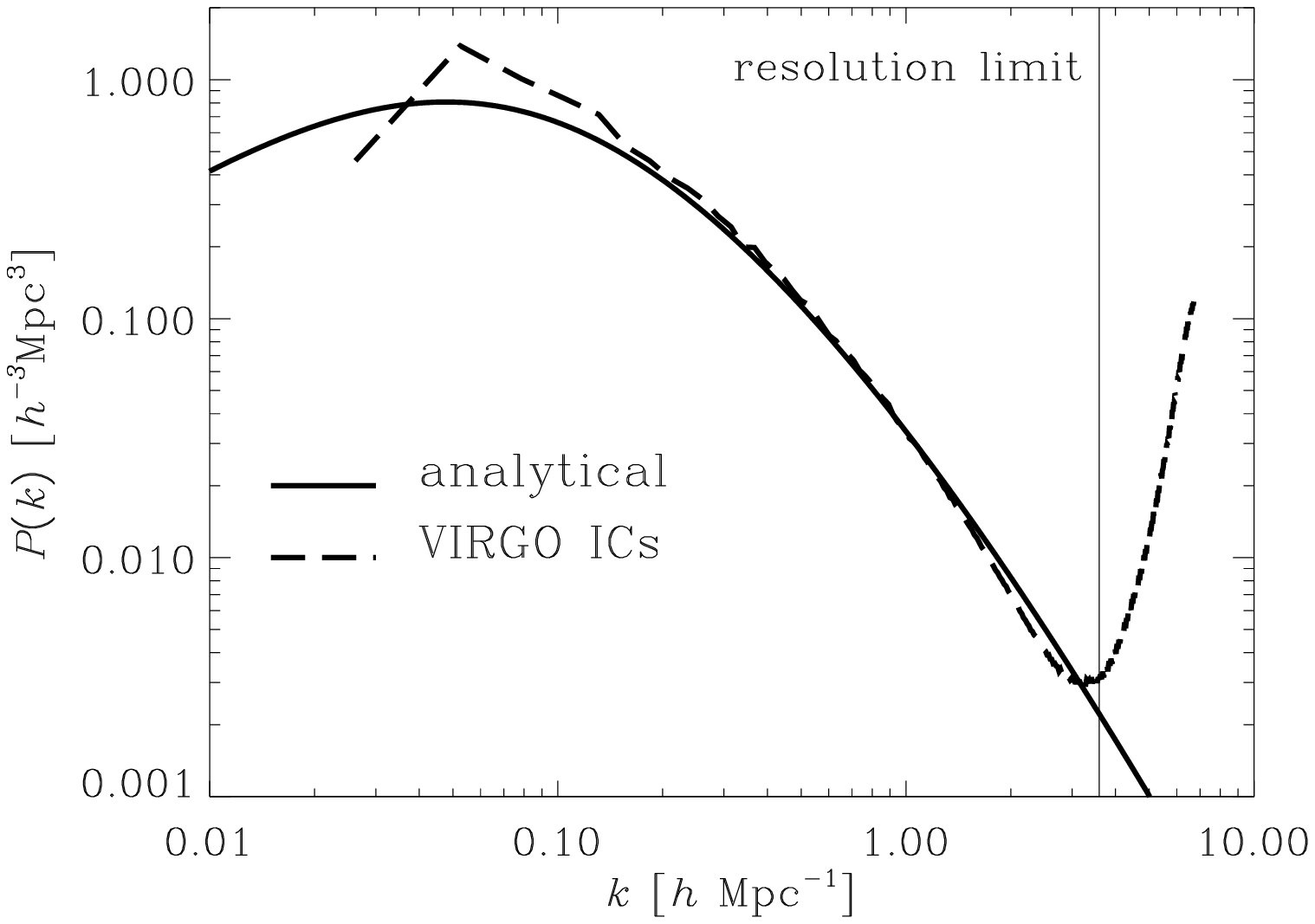}}
  \caption{
    {\it Left}: density variance in balls of radius $r$ (in units of
    the box size), estimated with our code and the one provided by the
    authors of \lett. {\it Right}: estimated power spectrum.  The
    resolution limit corresponds to a distance $\approx 2 \langle
    \Lambda \rangle$.}
  \label{fig:sigpow}
\end{figure}

\begin{figure}[h]
  \resizebox{.5\hsize}{!}{\includegraphics{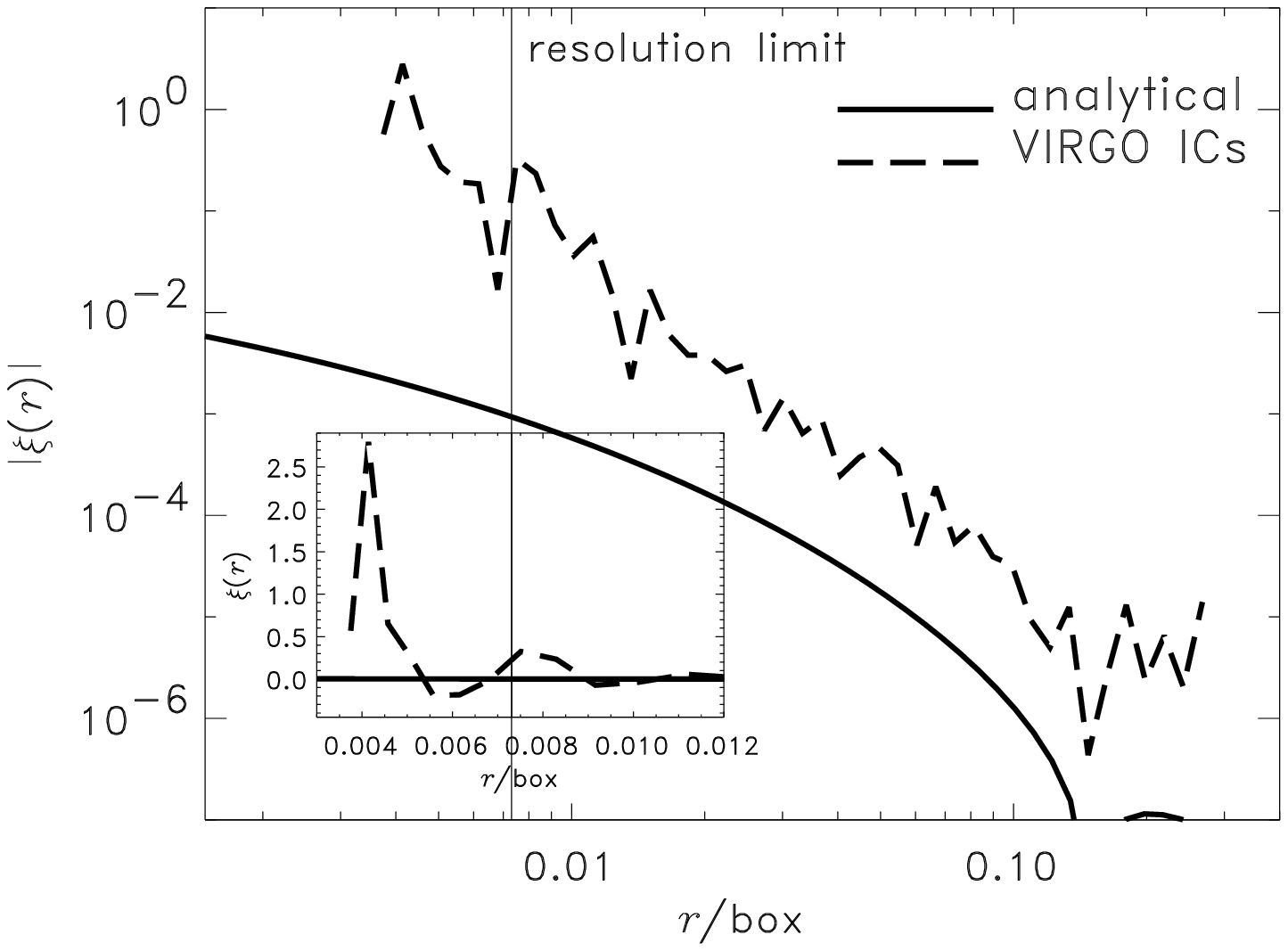}}
  \resizebox{.5\hsize}{!}{\includegraphics{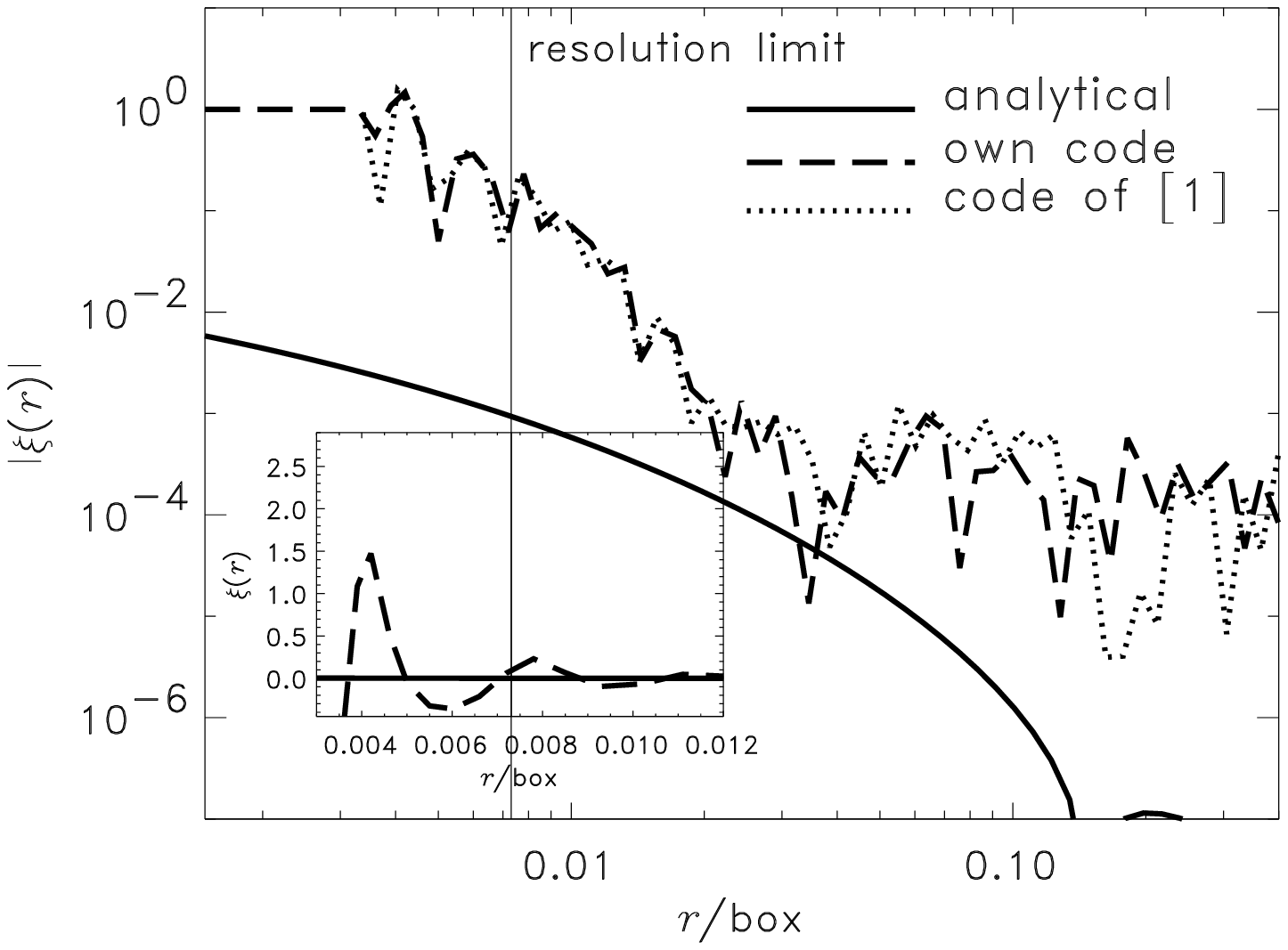}}
  \caption{
    Absolute value of the estimated correlation function; the inset
    panel shows the oscillatory behaviour of $\xi(r)$ for small
    separations. {\it Left}: random balls are allowed to overlap. {\it
      Right}: they are constrained to avoid overlap.}
  \label{fig:xi}
\end{figure}

Fig.~\ref{fig:sigpow} shows the density variance, $\sigma^2(r)$,
computed for the initial conditions of the SCDM1 model of the {\it Virgo Consortium} \cite{JFPT98}.
For $\sigma^2(r)$ we used the estimator quoted in \lett: variance of
density fluctuations in $20000$ spheres centered at random positions.
The theoretical CDM behaviors have been computed with the power
spectrum employed by the {\it Virgo Consortium} \cite{JFPT98}, and
they correct the wrong ones in \lett. For completeness, we also show
the estimated power spectrum, $P(k)$. We find that the estimated value
of $\sigma^2(r)$ {\em does follow} the theoretical CDM prediction in
an intermediate range of scales $2\langle \Lambda \rangle \lesssim r
\lesssim L/2$ ($\langle \Lambda \rangle$ being the mean interparticle
separation and $L$ the box size). On scales outside this trustworthy
range, the expected systematic departures due to discreteness and
finite-size set in. In the same way, the estimate of $P(k)$ also
agrees with the desired CDM behavior in the intermediate range $2\pi/L
\lesssim k \lesssim \pi/\langle \Lambda \rangle$.

Fig.~\ref{fig:xi} shows the estimated $\xi(r)$. We used the estimator
quoted in \lett, in turn a simple modification of the one for
$\sigma^2$ (variance in spherical {\em shells} centered at randomly
chosen {\em particles}).  The estimate of $\xi(r)$ exhibits noticeable
fluctuations around zero, whose amplitude is larger than the desired
CDM decay. This amplitude is comparable to the variation of the
estimate when using different random seeds for the random sphere
centers. Such a noise also arises when estimating $\xi(r)$ for simple
test cases whose correlation is known theoretically (uncorrelated
particles, particles at the nodes of a lattice). This suggests that
these fluctuations are mainly noise of the
$\xi$--estimator employed in \lett.  In these circumstances, a
systematic departure from the CDM behavior provoked by the method how
the IC's are generated cannot be assessed.  

We also learnt that the results shown in \lett\ were derived in
reality with a modification of the quoted estimators: the probing
spheres were constrained to avoid mutual overlap, so that their number
is effectively smaller than the intended 20000 for $r$ large enough.
We have observed that this leads to no detectable significant
difference when estimating $\sigma$, but the estimated $\xi (r)$ tends
to be larger in the large-$r$ end compared to the unconstrained
estimate.  This sensitivity on the number of probing spheres throws
further doubts on the reliability of the measured $\xi(r)$, and
it suggests looking for better estimators.

Our conclusions are: (i) $\sigma^2(r)$ and $P(k)$ reproduce very well
the desired CDM behavior in the expected intermediate range of scales;
(ii) the estimator of $\xi(r)$ employed in \lett\ is too noisy to
allow one to claim that $\xi(r)$ does or does not follow the expected
behavior; (iii) hence, these measurements provide no evidence against
the procedure employed to generate the IC's. 
%On the contrary, they support its reliability.

\end{document}